\documentclass[%
 aip,
cp,  
 amsmath,amssymb,
 reprint,%
]{revtex4-2}

\usepackage{graphicx}
\usepackage{dcolumn}
\usepackage{bm}

\usepackage[utf8]{inputenc}
\usepackage[T1]{fontenc}
\usepackage{mathptmx}

\begin{document}

\title{Elliptic Integrals for Calculation of the Propagation Time of a
Signal, Emitted and Received by Satellites on One Orbit}

\author{Bogdan G. Dimitrov} 
 \email[Corresponding author: ]{dimitrov.bogdan.bogdan@gmail.com}
\affiliation{
  Institute of Nuclear Research and Nuclear Energetics,
Bulgarian Academy of Sciences, 72 Tzarigradsko shaussee, 1784 Sofia,
Bulgaria.
}

\affiliation{%
Institute for Advanced Physical Studies , Sofia TEX Park, 111
Tzarigradsko shaussee, 1784 Sofia, Bulgaria  \\e-mail:
bogdan.dimitrov@iaps.institute}


\date{\today} 

\begin{abstract}
{\itshape} The propagation time of a signal, emitted by a moving along an
elliptical orbit satellite from the GPS (or GLONASS) satellite confi
gurations is a very important ingredient of the theory, based on the
formalism of the null cone and accounting for the effects of the General
Relativity Theory. For the case of satellites, orbiting along a plane
elliptic orbit, it has been proved that the propagation time for the signal between the satellites is given by a
combination of elliptic integrals of the first, second and third kind. For
the more general case of satellites on a space-distributed elliptic orbit,
the propagation time is expressed by higher (fourth) order elliptic
integrals, which according to the standard theory can be expressed
recurrently by means of lower-order elliptic integrals. In the concrete
case, the elliptic integrals of the second and the fourth order are
expressed by means of a combination of irrational functions and the
zero-order elliptic integral in the Legendre form. It has been proved that
for the investigated case, second-order elliptic integrals can be expressed
by elementary functions.
\end{abstract}

\maketitle

\section{\label{sec:Intro}Introduction}

In the past $15-20$ years the problem about GPS satellite-ground station
communications has been replaced by the problem about autonomous navigation
and intersatellite communications (ISC) (links), which has been mentioned
yet in 2005 in the monograph \cite{Moritz:C27}. Autonomous navigation means
that satellites on one orbit or on different orbits around the Earth should
have the capability to transmit data between them via intersatellite
cross-link ranging \cite{Xie:G1} and thus, to ensure navigation control and
data processing without commands from Earth stations in the course of six
months. However, since the transmitted signals are propagating in the space
around the Earth and are thus experiencing the influence of the
gravitational field of the Earth, their propagation should take into account
the General Relativity effects.

The theory of intersatellite communications (ISC) is developed in the series
of papers by S. Turyshev, V. Toth, M. Sazhin \cite{Turysh1:AAB62}, \cite%
{Turysh2:AAB63} and S. Turyshev, N. Yu, V. Toth \cite{Turysh3:AAB64}. This
theory concerns the space missions GRAIL (Gravity Recovery and Interior
Laboratory), GRACE- FOLLOW-ON (GRACE-FO - Gravity Recovery and Climate
Experiment - Follow On) mission and \ the Atomic Clock Ensemble in Space
(ACES) \ experiment on the International Space Station (ISS). However, it
should be stressed that all these missions are realized by means of
low-orbit satellites (about $450$ $km$ above the Earth), while the
satellites from the GPS, GLONASS and the Galileo constellations are on much
higher orbits (from $19140$ $km$ for GLONASS and ranging to $26560$ $km$ for
GPS, even higher). Consequently, since the main effect of the gravitational
field is to curve the trajectory of the signal and thus to increase the
propagation time of the signal and also the distance, travelled by the
signal (compared to the case of a flat spacetime), this effect will be
considerable for such large distances.

In the above-mentioned papers \cite{Turysh1:AAB62}, \cite{Turysh2:AAB63} and
\cite{Turysh3:AAB64}, the basic theoretical instrument for calculating the
propagation time of the signal is the Shapiro delay formulae. If the
coordinates of the emitting and of the receiving satellite are
correspondingly $\mid x_{A}(t_{A})\mid =r_{A}$ and $\mid x_{B}(t_{B})\mid
=r_{B}$, and $R_{AB}=$ $\mid x_{A}(t_{A})-x_{B}(t_{B})\mid $ is the
Euclidean distance between the signal - emitting satellite and the signal -
receiving satellite, then from the null cone equation, the signal
propagation time $T_{AB}=T_{B}-T_{A}$ between two space points can be
expressed by the known formulae \cite{Petit:AAB21} (see also the review
article \cite{Interf:CA7A2B1} by Sovers, Fanselow and Jacobs on VLBI radio
interferometry)
\begin{equation}
T_{AB}=\frac{R_{AB}}{c}+\frac{2GM_{E}}{c^{3}}\ln \left( \frac{r_{A}\nonumber%
\newline
+r_{B}+R_{AB}}{r_{A}+r_{B}-R_{AB}}\right) \ \ ,  \label{AA19}
\end{equation}%
where $G_{\oplus }M_{\oplus }$ is the geocentric gravitational constant, $%
M_{\oplus }$ is the Earth mass and $G_{\oplus }$ is the gravitational
constant. We shall denote also by $t$=TCG the Geocentric Coordinate Time
(TCG). The second term in formulae Eq. (\ref{AA19}) is the Shapiro time
delay term, \ accounting for the signal delay due to the curved space-time.
However, in this general form the Shapiro delay formulae cannot account for
the effect of the ellipticity of the orbit on the propagation of the signal
and also, the initial and the final radius-vectors $r_{A}$ and $r_{B}$
(related to the initial point of emission and the final point of reception
of the signal) should be dependent on some parameter, uniquely determining
the position of the satellite on the orbit. For the case of a plane
elliptical orbit this parameter will be the eccentric anomaly angle $E$ and
for space distributed orbits (parametrized by the full number of $6$
Keplerian elements $(f,a,e,\Omega ,I,\omega )$, the parameter is the true
anomaly angle $f$. In other words, in this paper the investigated problem
will be about propagation of a signal between satellites on one and the same
orbit.

The main goal of this paper is to find the explicite formulaes for the
propagation time of the signal for the two relevent cases - signal emitted
by a satellite, orbiting a plane orbit (and percepted by a satellite on the
same orbit), and signal, emitted by a satellite along a space-distributed
orbit (again, the signal-receiving satellite is on the same orbit). Although
in both cases the dynamical parameter will be just one (however, the
parameters will be different), the second case will be characterized by more
complicated formulaes, related to elliptical integrals of higher order. So
the general calculation should not be based on any direct application of the
Shapiro delay formulae Eq. (\ref{AA19}), but rather than that, on a
calculation, starting from the light null cone formalism.

There is a second motivation for finding the propagation time for the
various cases, related to the formalism of the two intersecting null cones
with origins at the signal-emitting and signal receiving satellites,
developed in the paper \cite{BOGDAN:PRD} (see also a shorter version in \cite%
{BOGDAN:AIP2019}). This formalism enables to find the propagation time of
the signal, when both satellites are moving, meaning that the propagation
time of the signal takes into account the motion of the second
(signal-receiving) satellite \ during the time for propagation of the signal
between the two satellites. However, the first ingredient of such a theory
is to find the s.c. "first propagation time", which is the propagation time
for the signal, emitted by the first satellite. In these papers, the theory
is developed only for the plane elliptical orbit case. In this paper, the
calculation for the first time is extended to the case of signal-emitting
and signal-receiving satellites on a space-oriented orbit. This might be
considered as the second step towards constructing a theory for propagation
of signals between moving satellites on different, space-oriented orbits.
The main motivation for such a complicated research (both from a physical
point of view and especially from a mathematical one) comes from the
requirement that the Global Navigation Satellite System ($GNSS$), consisting
of $30$ satellites on space-distributed orbits and orbiting the Earth at a
height of $23616$ km, should be interoperable with the other two
navigational systems $GPS$ and $GLONASS$ \cite{Soffel:C25}. This is
important from an operational point of view, since a combined GNSS of $75$
satellites from the GPS, GLONASS and the Galileo constellations may increase
greatly the visibility of the satellites, especially in critical areas such
as urban canyons \cite{Xu:C26}.

\section{\label{sec:PropPlaneOrbit}Propagation time for the case of a
signal-emitting satellite, moving along a plane elliptical orbit}

\subsection{\label{sec:PropWithoutAppr}Propagation time without any
approximations}

\bigskip Let us take the null-cone metric in the standard form
\begin{equation}
ds^{2}=-c^{2}\left( 1+\frac{2V}{c^{2}}\right) (dT)^{2}+\left( 1-\frac{2V_{.}%
}{c^{2}}\right) \left( (dx)^{2}+(dy)^{2}+(dz)^{2}\right) =0\text{ \ \ ,}
\label{DOP25}
\end{equation}%
where $V=\frac{G_{\oplus }M_{\oplus }}{r}$ is the standard gravitational
potential of the Earth. No account is taken of any any harmonics due to the
spherical form of the Earth since the GPS orbits are situated at a distance
more than $20000$ $km$ (considered from the centre of the Earth). Further,
the Kepler parametrization of the space coordinates $x=x(a,e,E)$ and $%
y=y(a,e,E)$ for the plane elliptical satellite orbit
\begin{equation}
x=a(\cos E-e)\text{ \ \ , \ \ \ \ }y=a\sqrt[.]{1-e^{2}}\sin E\text{ \ \ }
\label{ABCDF}
\end{equation}%
is used, where $a$ is the semi-major axis of the orbit, $e$ is the
eccentricity and $E$ is the eccentric anomaly angle, related to the motion
of the satellite along the plane elliptical orbit.

If the right-hand side of Eq. (\ref{DOP25}) is divided and multiplied by $%
(dt)^{2}$, where $t=t_{cel}$ is the celestial time from the Kepler equation $%
E-e\sin E=n(t_{cel}-t_{p})$ ($t_{p}$ is the time of perigee passage), one
can obtain
\begin{equation}
(c^{2}+2V)\left( dT\right) ^{2}=(1-\frac{2V}{c^{2}})\mathbf{v}^{2}\text{ \ ,}
\label{C5A25}
\end{equation}%
where $\mathbf{v}^{2}$ is the square of the satellite velocity along the
orbit
\begin{equation}
\mathbf{v}^{2}=v_{x}^{2}+v_{y}^{2}=\left( \frac{dx}{dt}\right) ^{2}+\left(
\frac{dy}{dt}\right) ^{2}\text{ \ .}  \label{C5A26}
\end{equation}%
Taking into account the Kepler parametrization Eq. (\ref{ABCDF}), the above
expression for the velocity for the case of plane motion can be rewritten as
\begin{equation}
\mathbf{v=}\sqrt{v_{x}^{2}+v_{y}^{2}}=\frac{na}{\left( 1-e\cos E\right) }%
\sqrt[.]{1-e^{2}\cos ^{2}E}\text{ \ \ }  \label{C5A27}
\end{equation}%
and $n=\sqrt[.]{\frac{G_{\oplus }M_{\oplus }}{a^{3}}}$ is the mean motion.
Expressing $dT$ from Eq. (\ref{C5A25}), finding the celestial time $t_{cel}$
in terms of $E$ from the Kepler equation and performing the integration over
the eccentric anomaly angle $E$, one can obtain the expression for the
propagation time (for the case of no restrictions imposed)

\begin{equation}
T=\int \frac{\mathbf{v}}{c}\sqrt[.]{\frac{(c^{2}-2V)}{(c^{2}+2V)}}\text{ }%
dE+C=\frac{a}{c}\int \sqrt[.]{\frac{a_{1}y^{3}+a_{2}y^{2}+a_{3}y}{%
b_{1}y^{3}+b_{2}y^{2}+b_{3}y+b_{4}}}dy\text{ \ \ \ ,}  \label{C5A33}
\end{equation}%
where the numerical constants $a_{1}$, $a_{2}$, $a_{3}$, $b_{1}$, $b_{2}$, $%
b_{3}$, $b_{4}$ can easily be calculated, $y$ is the variable $y=1-e\cos E$
and $C$ is some integration constant, which can be assumed to be zero.
Integrals of the type Eq. (\ref{C5A33}) are not abelian ones (see the
monograph by Prasolov and Solovyev \cite{PrasSol}), because abelian
integrals are related to algebraic curves $F(x,y):=x^{2}-P(y)=0$, where $%
P(y) $ is an algebraic polynomial. In the case, the integrand 
expression is a rational function of $y$. Further the expression Eq. (\ref{C5A33})
will not be used, because a justifiable and reasonable physical
approximation shall be applied. Nevertheless, in view of the constantly
improving accuracy for measuring the propagation time, this might represent
an interesting problem for future mathematical research.

\subsection{\label{sec:PlaneWithApprox}The useful approximation for a small
gravitational potential, compared to the square of the velocity of \ light}

Further we shall be interested in the case
\begin{equation}
\beta =\frac{2V}{c^{2}}=\frac{2G_{\oplus }M_{\oplus }}{c^{2}a}\ll 1\text{ \
\ , }  \label{C5A37}
\end{equation}%
which can \ be assumed to be fulfilled. For the parameters of the $GPS$
orbit - $a=26561$ $[km]$ , in the review paper \cite{Sanchez} the constant $%
\beta $ can exactly be calculated to be $0.334\times 10^{-9}$, with the
velocity of light taken to be $c=299792458$\ $[\frac{m}{\sec }].$\  
  The geocentric gravitational constant $G_{\oplus }M_{\oplus }$ (obtained from
the analysis of laser distance measurements of artificial Earth satellites)
can be taken to be equal to $G_{\oplus }M_{\oplus }=\left( 3986004.405\pm
1\right) \times 10^{8}\ [\frac{m^{3}}{\sec ^{2}}]$, but the value of $%
G_{\oplus }M_{\oplus }$ can vary also in another range from $G_{\oplus
}M_{\oplus }=3986056.75236\times 10^{8}\ [\frac{m^{3}}{\sec ^{2}}]$ to the
value $G_{\oplus }M_{\oplus }=3987999.07898\times 10^{8}\ [\frac{m^{3}}{\sec
^{2}}]$ due to the uncertainties in measuring the Newton gravitational
constant $G_{\oplus }$. The mass of the Earth can be taken approximately to
be $M_{E}\approx 5.97\times 10^{24}$ $[kg]$. One of the latest values for $%
G_{\oplus }$ from deep space experiments was reported in the paper \cite%
{KopMuller2019} to be $(6.674+0.0003).10^{-11}$ $[\frac{m^{3}}{kg.\sec ^{2}}%
] $.

\subsection{\label{sec:PhysJustif}Derivation of the propagation time under
the approximation $\protect\beta =\frac{2V}{c^{2}}\ll 1$ and physical
justification of the obtained result}

After decomposing the under-integral expression with the square root in Eq. (%
\ref{C5A33}) and leaving only the first-order term in $\frac{2V}{c^{2}}$,
one can obtain
\begin{equation}
T=\int \frac{\mathbf{v}}{c}\sqrt[.]{\frac{(1-\frac{2V}{c^{2}})}{(1+\frac{2V}{%
c^{2}})}}\text{ }dt\approx \int \frac{\mathbf{v}}{c}(1-\frac{2V}{c^{2}}%
)dt=I_{1}+I_{2}=  \label{C5A39}
\end{equation}%
\begin{equation}
=\frac{a}{c}\int \sqrt[.]{1-e^{2}\cos ^{2}E}\text{ }dE-\frac{2G_{\oplus
}M_{\oplus }}{c^{3}}\int \sqrt[.]{\frac{1+e\cos E}{1-e\cos E}}dE\text{ \ \ .}
\label{C5A40}
\end{equation}%
This expression is consistent from a physical point of view due to the
following reasons:

1. The coefficient $\frac{a}{c}$ as a ratio of the large semi-major axis of
the orbit and the velocity of light $c=299792458$ \ $[\frac{m}{\sec }]$ will
have a dimension $[m/\frac{m}{\sec }]=[\sec ]$, as it should be. The second
coefficient $\frac{2G_{\oplus }M_{\oplus }}{c^{3}}$ has a corresponding
dimension $[\frac{m^{3}}{\sec ^{2}}:\frac{m^{3}}{\sec ^{3}}]=[\sec ]$, which
clearly proves that formulae Eq. (\ref{C5A40}) has the proper dimensions.

2. The formulae in fact gives the propagation time $T$ of the signal,
emitted by the satellite at some initial position (given by the eccentric
anomaly angle $E_{init}$), and the final point (given by $E_{fin}$) of \
reception of the signal by another satellite. So both emission- and
reception- signal points should remain on one and the same satellite orbit.
In this sense, the integrals Eq. (\ref{C5A39}) and Eq. (\ref{C5A40})
represent curvilinear integrals (depending on the initial and the final
points of integration), derived after the application of the null-cone
equation (in a general form written as $F=g_{\alpha \beta }dx^{\alpha
}dx^{\beta }=0$) and consequently, they are related to the propagation time
of the signal and not to the motion of the satellite. Moreover, the
condition $F=0$ is proved to be compatible with the geodesic equation \cite%
{Fock} for light-like geodesics$\ \ \ \ \ $
\begin{equation}
\frac{d^{2}x_{\nu }}{d\tau ^{2}}+\Gamma _{\alpha \beta }^{\nu }\frac{%
dx^{\alpha }}{d\tau }\frac{dx^{\beta }}{d\tau }=0\text{ \ \ \ ,}
\label{C5A41}
\end{equation}%
where $\tau $ is the proper time along the light ray and $\nu ,\alpha ,\beta
=0,1,2,3$. In other words, the zero-length (light-like) geodesics
(representing the trajectory of the signal) are determined also by the null
cone equation $g_{\alpha \beta }dx^{\alpha }dx^{\beta }=0$, because the null
equation is a first integral of the geodesic equation Eq. (\ref{C5A41}).

3. Let now $x,y,z$ be the space coordinates, reached by the signal after it
has been emitted. From the null cone equation Eq. (\ref{DOP25}), after using
the approximation $\beta =\frac{2V}{c^{2}}\ll 1$, it can be obtained
\begin{equation}
\left( (dx)^{2}+(dy)^{2}+(dz)^{2}\right) \approx c^{2}\left[ 1+\frac{4V}{%
c^{2}}\right] (dT)^{2}\ll 3c^{2}(dT)^{2}\text{ \ .}  \label{CC5A41}
\end{equation}%
Since $cdT$ is the infinitesimal distance, travelled by the light signal,
the above inequality means that the distance travelled by light is much
greater than the Euclidean distance. In other words, the light trajectory
signal is a curved one, in accord with the meaning of the Shapiro delay
formulae Eq. (\ref{AA19}) about the delay of the signal under the action of
the gravitational field. \

4. The trigonometric function $\cos ^{2}E$ has the same values for $E=\alpha
$ and $E=180-\alpha $, so the correspondence eccentric anomaly angle $E$ and
the propagation time $T$ is not reliable for angles in the second quadrant.

5. The propagation time $T$ Eq. (\ref{C5A40}) should be a real-valued
expression, since time cannot be complex. This fact is not evident from the
beginning of the calculations, but will be proved in the next subsections.
Remarkably, this important property will turn out to be valid also for the
next case of a signal, emitted and percepted by satellites on a
space-distributed orbit.

\subsection{\label{sec:VariousElliptic}Mathematical structure of the
expression for the propagation time, related to zero-order elliptic
integrals of the first, second and the third rank}

The first term in expression Eq. (\ref{C5A40}) after performing the simple
transformation $\frac{\pi }{2}-E=\overline{E}$ can be brought to the
following form
\begin{equation}
T_{1}=\int\limits_{0}^{E}\sqrt[.]{1-e^{2}\cos ^{2}E}dE=\int\limits_{0}^{E}%
\sqrt[.]{1-e^{2}\sin ^{2}(\frac{\pi }{2}-E)}dE=-\int\limits_{\frac{\pi }{2}%
}^{\frac{\pi }{2}-\overline{E}}\sqrt[.]{1-e^{2}\sin ^{2}\overline{E}}d%
\overline{E}\text{ \ .}  \label{C30}
\end{equation}%
This represents an elliptic integral of the second kind.

The second term in Eq. (\ref{C5A40}) after performing the substitution
\begin{equation}
\sqrt[.]{\frac{1+e\cos E}{1-e\cos E}}=\overline{y}  \label{EE1}
\end{equation}%
and introducing the notations
\begin{equation}
\widetilde{k}^{2}=\frac{1-e}{1+e}=q\text{ \ \ , \ }\frac{\overline{y}}{%
\widetilde{k}}=y\text{ \ \ \ , \ \ }\left( \frac{\overline{y}}{\widetilde{k}}%
\right) ^{2}=\widetilde{y}\text{ \ \ , \ }\widetilde{y}=-\widetilde{%
\widetilde{y}}  \label{E8}
\end{equation}%
can be written as a sum of two integrals, i.e.
\begin{equation}
T_{2}=-\frac{2G_{\oplus }M_{\oplus }}{c^{3}}\int \sqrt[.]{\frac{1+e\cos E}{%
1-e\cos E}}dE=I_{2}^{A)}+I_{2}^{(B)}\text{ \ .}  \label{E9}
\end{equation}%
It should be noted that $\widetilde{k}$ in eq. (\ref{E8}) is merely a
notation, representing the real expression $\widetilde{k}=\sqrt{\frac{1-e}{%
1+e}}=\sqrt{q}$, since $0<\frac{1-e}{1+e}<1$ (the eccentricity of the
orbit $e$ is always less than one,  \ $0<e<1$). Using this notation, the
first term $I_{2}^{(A)}$ in eq.  (\ref{E9}) can be represented as
\begin{equation}
I_{2}^{(A)}=\frac{4GM}{c^{3}}\frac{1}{\widetilde{k}\sqrt[.]{e^{2}-1}}\int
\frac{d\widetilde{y}}{\sqrt[.]{\widetilde{y}\left( \widetilde{y}-\frac{1}{%
\widetilde{k}^{4}}\right) \left( \widetilde{y}-1\right) }}\text{ \ \ \ .}
\label{E10A2}
\end{equation}%
Due to the presence of the $\sqrt[.]{e^{2}-1}$ term in the denominator, it
might seem that expression Eq. (\ref{E10A2}) is imaginary. However, this is
not true, because it can be rewritten as
\begin{equation}
I_{2}^{(A)}=\frac{4GM}{c^{3}}\frac{1}{\widetilde{k}\text{ }i\sqrt[.]{1-e^{2}.%
}\text{ }\left( -i\right) }\int \frac{d\widetilde{\widetilde{y}}}{\sqrt[.]{%
\text{ }\widetilde{\widetilde{y}}\left( \widetilde{\widetilde{y}}+1\right)
\left( \widetilde{\widetilde{y}}+\frac{1}{\widetilde{k}^{4}}\right) }}\text{
\ \ ,}  \label{E10A4}
\end{equation}%
representing an elliptic integral of zero order and of the first kind. Since
$\widetilde{k}$ is real-valued expression, the two imaginary units $i$ in
the denominator appear from $\sqrt[.]{e^{2}-1}=i\sqrt[.]{1-e^{2}}$ and also
from the square root, after performing the variable transformation from  $%
\widetilde{y}$ \ to \ $\widetilde{\widetilde{y}}$ in $\widetilde{y}\left(
\widetilde{y}-\frac{1}{\widetilde{k}^{4}}\right) \left( \widetilde{y}%
-1\right) $ in eq. (\ref{E10A2}). As a result, a factor $\sqrt{(-1)^{3}}=%
\sqrt{i^{6}}=i^{3}=-i$ will appear in the denominator of eq. (\ref{E10A4}).
The whole expression eq. (\ref{E10A4}) turns out to be a real-valued one and
with a negative sign, because $d\widetilde{y}=-d\widetilde{\widetilde{y}}$ .
Since this expression is a part of the expression for the propagation time,
this is physically reasonable.

The second term $I_{2}^{(B)}$ in Eq. (\ref{E9}) can also be written in the
form of a real-valued expression, as it should be
\begin{equation}
I_{2}^{(B)}=\frac{4GM}{c^{3}q^{2}}\frac{1}{\sqrt[.]{1-e^{2}}}\int \frac{d%
\widetilde{y}}{(\widetilde{y}-\frac{1}{q})\sqrt[.]{\widetilde{y}(\widetilde{y%
}+1)(\widetilde{y}+\frac{1}{q^{2}})}}\text{ \ .}  \label{E13}
\end{equation}%
This is an elliptic integral of the third kind. Consequently, the whole
expression for the propagation time can be represented as a sum of elliptic
integrals of the second, the first and the third kind \cite{PrasSol}.

\section{\label{sec:SpaceDistrOrbit}Propagation time for a signal, emitted
and percepted by satellites on a space-distributed orbit}

\subsection{\label{sec:3DParam}Three-dimensional orbit parametrization and
the general formulae for the orbit parametrization}

Instead of the simple two-dimensional parametrization Eq. (\ref{ABCDF}), the
three-dimensional parametrization of the space-distributed orbit to be used
is \cite{KopEfroim}
\begin{equation}
x=\frac{a(1-e^{2})}{1+e\cos f}\left[ \cos \Omega \cos (\omega +f)-\sin
\Omega \sin (\omega +f)\cos i\right] \text{ \ \ ,}  \label{K1}
\end{equation}%
\begin{equation}
y=\frac{a(1-e^{2})}{1+e\cos f}\left[ \sin \Omega \cos (\omega +f)+\cos
\Omega \sin (\omega +f)\cos i\right] \text{ \ \ ,}  \label{K2}
\end{equation}%
\begin{equation}
z=\frac{a(1-e^{2})}{1+e\cos f}\sin (\omega +f)\sin i\text{ \ \ \ ,}
\label{K3}
\end{equation}%
where $r=\frac{a(1-e^{2})}{1+e\cos f}$ \ is the radius-vector in the orbital
plane, the angle $\Omega $\ of the longitude of the right ascension of the
ascending node is the angle between the line of nodes and the direction to
the vernal equinox, the argument of perigee (periapsis) $\omega $ is the
angle within the orbital plane from the ascending node to perigee in the
direction of the satellite motion $(0\leq \omega \leq 360^{0})$, the angle $%
i $ is the inclination of the orbit with respect to the equatorial plane and
the true anomaly $f$ geometrically represents the angle between the line of
nodes and the position vector $r$ on the orbital plane has an initial point
at the centre of the ellipse. Since the angle $f$ is related to the motion
of the satellite and all the other parameters of the orbit do not change
during the motion of the satellite, it can easily be found that \textit{\ }
\begin{equation}
\sqrt{(dx)^{2}+(dy)^{2}+(dz)^{2}}=\sqrt{%
(v_{f}^{x})^{2}+(v_{f}^{y})^{2}+(v_{f}^{z})^{2}}df=v_{f}df\text{ \ \ ,}
\label{KK4}
\end{equation}%
\ where the velocity $v_{f}$\ (with velocity components $%
(v_{f}^{x},v_{f}^{y},v_{f}^{z}$), associated to the true anomaly angle $f$
is given by
\begin{equation}
v_{f}=\frac{na}{\sqrt{1-e^{2}}}\sqrt{1+e^{2}+2e\cos f}\text{ \ . }
\label{K4}
\end{equation}%
Further, again making use of the null cone equation Eq. (\ref{C5A25}) and
also of the approximation Eq. (\ref{C5A37}) $\beta =\frac{2V}{c^{2}}=\frac{%
2G_{\oplus }M_{\oplus }}{c^{2}a}\ll 1$, one can obtain the general formulae
for the propagation time
\begin{equation}
T=\int \frac{v}{c}(1-\frac{2V}{c^{2}})dt=\widetilde{T}_{1}+\widetilde{T}_{2}=%
\frac{1}{c}\int vdt-\frac{2}{c^{3}}\int vVdt\text{ \ \ .}  \label{K5}
\end{equation}%
Note the important fact that in the above formulae one can replace $vdt$ by $%
v_{f}df$, because
\begin{equation}
vdt=\sqrt{\left( \frac{dx}{dt}\right) ^{2}+\left( \frac{dy}{dt}\right)
^{2}+\left( \frac{dz}{dt}\right) ^{2}}dt=\sqrt{\left( \frac{dx}{df}\right)
^{2}+\left( \frac{dy}{df}\right) ^{2}+\left( \frac{dz}{df}\right) ^{2}}%
df=v_{f}df\text{ .}  \label{K6}
\end{equation}%
It is worth mentioning that without the approximation Eq. (\ref{C5A37}), a
similar formulae to Eq. (\ref{C5A33}) can be obtained. Here we shall not
write this formulae.

\subsection{\label{sec:3DAnalWithoutEllip}Analytical calculation of the
first integral (first $O(\frac{1}{c})$) correction without the use of
elliptic integrals}

We shall present a calculation, which shows that some integrals can be
calculated both analytically, and also by the use of elliptic integrals.

Let us take the first $O(\frac{1}{c})$ propagation time correction
\begin{equation}
\widetilde{T}_{1}=\frac{1}{c}\int v_{f}df=\frac{na}{c\sqrt{1-e^{2}}}\int
\sqrt{1+e^{2}+2e\cos f}df=\frac{na}{c\sqrt{1-e^{2}}}\widetilde{T}_{1}^{(1)}
\label{K7}
\end{equation}

and let us perform the series of \ six subsequent variable transformations
\begin{equation}
\sqrt{1+e^{2}+2e\cos f}=y\text{ \ \ \ , \ \ \ }\overline{y}=(e+1)^{2}-y^{2}%
\text{ \ \ \ ,}  \label{K8}
\end{equation}%
\begin{equation}
\sqrt{1-\frac{\overline{y}q^{2}}{\overline{y}+B_{1}}}=z\text{ \ , \ \ \ \ }%
z^{2}=z_{1}\text{ \ \ , \ }1-z_{1}=m\text{ \ , \ \ }\overline{m}=m-\frac{%
q^{2}}{2}\text{\ .\ \ }  \label{KK8}
\end{equation}%
Then for $\widetilde{T}_{1}$ the following sum of two integrals will be
obtained \
\begin{equation}
\widetilde{T}_{1}=-\frac{nai}{c\sqrt{(1-e^{2})q}}\left[ \int \frac{d%
\overline{m}}{\left( \overline{m}-\frac{q^{2}}{2}\right) \sqrt{\overline{m}-%
\frac{q^{2}}{2}}}+\int \frac{d\overline{m}}{\left( \overline{m}+\frac{q^{2}}{%
2}\right) \sqrt{\overline{m}-\frac{q^{2}}{2}}}\right] \text{ \ , \ \ \ }
\label{K9}
\end{equation}%
where again the notation $q=\frac{1-e}{1+e}$ has been used and each of the
integrals inside the square bracket will be denoted correspondingly as $%
\widetilde{T}_{1}^{(1)}$ and $\widetilde{T}_{1}^{(2)}$. Making use of the
analytically calculated integral from the book \cite{Timof}
\begin{equation}
\int \frac{dx}{(x\pm p)\sqrt{a+2bx+cx^{2}}}=-\frac{1}{\sqrt{a\mp 2bp+cp^{2}}}%
\ln \left\{ \frac{\sqrt{a+2bx+cx^{2}}+\sqrt{a\mp 2bp+cp^{2}}}{x\pm p}+\frac{%
b\mp cp}{\sqrt{a\mp 2bp+cp^{2}}}\right\} \text{ ,}  \label{K10}
\end{equation}%
one can derive for the second integral $\widetilde{T}_{1}^{(2)}$ in Eq. (\ref%
{K9}) the following expression
\begin{equation}
\widetilde{T}_{1}^{(2)}=-\frac{na(1+e)\sqrt{2}}{c\sqrt{1-e^{2}}\sqrt{q}\sqrt{%
3e^{2}+2e+3}}\ln \left[ \left( \frac{\sqrt{2}m_{1}(f_{b};r_{b})+\frac{1}{2}%
m_{2}(f_{b};r_{b})}{\sqrt{2}m_{1}(f_{a};r_{a})+\frac{1}{2}m_{2}(f_{a};r_{a})}%
\right) \left( \frac{m_{3}(f_{a};r_{a})}{m3(f_{b};r_{b})}\right) \right]
\label{K11}
\end{equation}%
and $m_{1}(f;r)$, $m_{2}(f;r)$, $m_{3}(f;r)$ are expressions, written in
terms either of the initial and final true anomaly angles $f_{a}$ and $f_{b}$
or, of the initial distance $r_{a}$ (at which the emission of the signal
takes place) and the final point $r_{b}$ of reception on the same orbit,
corresponding to the propagation time $\widetilde{T}_{1}^{(2)}$. The first
integral $\widetilde{T}_{1}^{(1)}$ can be calculated analogously. It should
be noted that both $\widetilde{T}_{1}^{(1)}$ and $\widetilde{T}_{1}^{(2)}$
are real-valued expressions and moreover, the logarithmic term is again
present, as in the original Shapiro delay formulae Eq. (\ref{AA19}).

\subsection{\label{sec:3DAnalWithElliptic}Analytical calculation of the
first $O(\frac{1}{c})$ correction by means of elliptic
integrals}

Let us calculate the integral $\widetilde{T}_{1}=\frac{na}{c\sqrt{1-e^{2}}}%
\int \sqrt{1+e^{2}+2e\cos f}df$ \ by making use of the substitution
\begin{equation}
y=\sqrt{\frac{1}{q}.\frac{(1+e\cos E)}{(1-e\cos E)}}\text{ \ \ , \ }q=\frac{%
1-e}{1+e}  \label{K12}
\end{equation}%
and also of the well-known relation from celestial mechanics between the
eccentric anomaly angle $E$ and the true anomaly angle $f$ \ \cite{KopEfroim}
\begin{equation}
\tan \frac{f}{2}=\sqrt{\frac{1-\cos f}{1+\cos f}}=\sqrt{\frac{1+e}{1-e}}\tan
\frac{E}{2}\text{ \ \ \ .}  \label{K13}
\end{equation}%
Then we can write the integral $\widetilde{T}_{1}$\bigskip\ in the form of
an elliptic integral of the second order and of the first kind in the
Legendre form
\begin{equation}
\widetilde{T}_{1}=-2i\frac{na}{c}q^{\frac{3}{2}}\int \frac{y^{2}dy}{\sqrt{%
(1-y^{2})(1-q^{2}y^{2})}}=-2i\frac{na}{c}q^{\frac{3}{2}}I\text{ \ \ \ .}
\label{K14}
\end{equation}%
The integral $I$ can also be represented as
\begin{equation}
I=\widetilde{J}_{2}^{(4)}(y;q)=\int \frac{y^{2}dy}{\sqrt{%
(1-y^{2})(1-q^{2}y^{2})}}=-\frac{1}{q^{2}}\int \frac{(1-q^{2}y^{2})dy}{\sqrt{%
(1-y^{2})(1-q^{2}y^{2})}}+\frac{1}{q^{2}}\int \frac{dy}{\sqrt{%
(1-y^{2})(1-q^{2}y^{2})}}=  \label{K15}
\end{equation}%
\begin{equation}
=-\frac{1}{q^{2}}\int \sqrt{1-q^{2}\sin ^{2}\varphi }d\varphi +\frac{1}{q^{2}%
}\int \frac{dy}{\sqrt{(1-y^{2})(1-q^{2}y^{2})}}\text{ \ \ .}  \label{K16}
\end{equation}%
The first integral in Eq. (\ref{K16}) is an elliptic integral of the second
kind (denoted usually by $E(\varphi )=\int \sqrt{1-q^{2}\sin ^{2}\varphi }%
d\varphi $, where $x=\sin \varphi $). So we obtain a relation between the
second-order elliptic integral $\widetilde{J}_{2}^{(4)}(y;q)$ of the first
kind in the Legendre form, the zero-order elliptic integral of the first
kind in the Legendre form $\widetilde{J}_{0}^{(4)}(y;q)$ (the second
integral in Eq. (\ref{K16})) and the elliptic integral $E(\varphi )$
\begin{equation}
\widetilde{J}_{2}^{(4)}(y;q)=\int \frac{y^{2}dy}{\sqrt{%
(1-y^{2})(1-q^{2}y^{2})}}=-\frac{1}{q^{2}}E(\varphi )+\frac{1}{q^{2}}%
\widetilde{J}_{0}^{(4)}(y;q)\text{ \ \ \ \ .}  \label{K17}
\end{equation}%
Since in the preceding section an analytical expression was obtained for $%
\widetilde{T}_{1}$ without the use of any elliptic functions, the obtained
result in Eq. (\ref{K16}) means that for the investigated case, elliptic
integrals of second order can be expressed through elementary functions,
contrary to the claims of some authors that this is not possible for all
kinds of elliptic integrals.

Using the relation between the eccentric anomaly angle $E$ and the true
anomaly angle $f$ (derived from Eq. (\ref{K13}) ), it is interesting to
express the first $(O(\frac{1}{c}))$ propagation time correction Eq. (\ref%
{K14}) also in terms of the variable $E$
\begin{equation}
\widetilde{T}_{1}=\frac{na}{c}\sqrt{1-e^{2}}\int \frac{1}{(1-e\cos E)}\sqrt{%
\frac{1+e\cos E}{1-e\cos E}}dE\text{ \ \ .}  \label{KK17}
\end{equation}%
This integral resembles the second integral $-\frac{2G_{\oplus }M_{\oplus }}{%
c^{3}}\int \sqrt[.]{\frac{1+e\cos E}{1-e\cos E}}dE$ in the $O(\frac{1}{c^{3}}%
)$ time correction Eq. (\ref{C5A40}) for the case of plane elliptical orbit,
but in the case the integral is with another coefficient and is modified
with the term $\frac{1}{(1-e\cos E)}$, multiplying the square root.

\subsection{Second analytical calculation of the first time $O(\frac{1}{c})$
correction in terms of second-order elliptic integrals}

\bigskip This second calculation will not make any use of the eccentric
anomaly angle variable $E$. Let us apply the variable transformation
\begin{equation}
\widetilde{y}=\frac{\sqrt{1+2e\cos f+e^{2}}}{1+e}=\frac{\widetilde{y}}{1+e}%
\text{ \ ,}  \label{K18}
\end{equation}%
after which the integral $\widetilde{T}_{1}$ acquires the form
\begin{equation}
\widetilde{T}_{1}=\frac{na}{c\sqrt{1-e^{2}}}\int \sqrt{1+e^{2}+2e\cos f}df=
\label{K19}
\end{equation}%
\begin{equation}
=i\frac{2na(1+e)}{cq\sqrt{1-e^{2}}}\int \frac{\widetilde{y}^{2}d\widetilde{y}%
}{\sqrt{\left( 1-\widetilde{y}^{2}\right) \left( 1-\frac{\widetilde{y}^{2}}{%
q^{2}}\right) }}\text{ }=\frac{2na(1+e)}{cq\sqrt{1-e^{2}}}\int \frac{%
\widetilde{y}^{2}d\widetilde{y}}{\sqrt{\left( 1-\widetilde{y}^{2}\right)
\left( \frac{\widetilde{y}^{2}}{q^{2}}-1\right) }}\text{\ \ .}  \label{K20}
\end{equation}%
Note that the coefficient in front of the integral is modified in comparison
with the one in Eq. (\ref{K14}), and more importantly, the resulting
expression is again a real-valued one \ due to the property of the elliptic
integral
\begin{equation}
\widetilde{J}_{2}^{(4)}(\widetilde{y},\frac{1}{q})=\frac{1}{i}\int \frac{%
\widetilde{y}^{2}d\widetilde{y}}{\sqrt{\left( 1-\widetilde{y}^{2}\right)
\left( \frac{\widetilde{y}^{2}}{q^{2}}-1\right) }}\text{ \ \ \ .}
\label{K21}
\end{equation}%
This formulae is valid, because due to the inequality $\cos f\leq 1$ and the
choice of the variable $\widetilde{y}$ in Eq. (\ref{K18}), it can easily be
proved that
\begin{equation}
\widetilde{y}\leq 1\text{ \ , }\vartriangleright 1-\widetilde{y}^{2}\geq 0%
\text{ \ , \ \ }1-\frac{\widetilde{y}^{2}}{q^{2}}\geq -\frac{(1-q^{2})}{q^{2}%
}\text{ \ .}  \label{K22}
\end{equation}%
Consequently, since $1-\frac{\widetilde{y}^{2}}{q^{2}}$ can take negative
values (note that $q^{2}=\left( \frac{1-e}{1+e}\right) ^{2}<1$), the
representation Eq. (\ref{K21}) is correct.

\section{\label{sec:HigherOrderCompar}Definitions of elliptic integrals of
higher order and comparison with some statements from standard textbooks}

\subsection{\protect\bigskip \label{sec:HigherOrder}Higher-order elliptic
integrals}

This definition is necessary to be given because further, when calculating
the second part of the propagation time, we shall encounter elliptic
integrals of the fourth-order.

According to the general definition, elliptic integrals are of the type
\begin{equation}
\int R(y,\sqrt{G(y)})dy\text{ \ or}\int R(x,\sqrt{P(x)})dx\text{ \ \ ,}
\label{II.1}
\end{equation}%
where $R(\widetilde{y},y)$ and $R(\widetilde{x},x)$ are rational functions
of the variables $\widetilde{y},y$ or $\widetilde{x},x$ and $\widetilde{y}=%
\sqrt{G(y)}$, $\widetilde{x}=\sqrt{P(x)}$ are arbitrary polynomials of the
fourth or of the third degree respectively. In accord with the notations in
the monograph \cite{PrasSol}, the fourth-order and the third-order
polynomials will be of the form
\begin{equation}
\widetilde{y}^{2}=G(y)=a_{0}y^{4}+4a_{1}y^{3}+6a_{2}y^{2}+4a_{3}y+a_{4}\text{
\ \ , \ }\widetilde{x}^{2}=P(x)=ax^{3}+bx^{2}+cx+d\text{ \ .\ }  \label{II.2}
\end{equation}%
So in the general case, elliptic integrals of the $n-$th order and of the
first kind are defined as
\begin{equation}
J_{n}^{(4)}=\int \frac{y^{n}dy}{\sqrt{%
a_{0}y^{4}+4a_{1}y^{3}+6a_{2}y^{2}+4a_{3}y+a_{4}}}\text{ \ , \ }%
J_{n}^{(3)}=\int \frac{x^{n}dx}{\sqrt{ax^{3}+bx^{2}+cx+d\text{ }}}
\label{II.11A}
\end{equation}%
Elliptic integrals of the $n-$th order and of the third kind are defined as
\begin{equation}
H_{n}^{(4)}=\int \frac{dy}{(y-c_{1})^{n}\sqrt{%
a_{0}y^{4}+4a_{1}y^{3}+6a_{2}y^{2}+4a_{3}y+a_{4}}}\text{ \ , \ }%
H_{n}^{(3)}=\int \frac{dx}{(x-c_{2})^{n}\sqrt{ax^{3}+bx^{2}+cx+d\text{ }}}%
\text{ \ .}  \label{II.11B}
\end{equation}%
In the above definitions, the upper indices "$(3)$" or "$(4)$" denote the
order of the algebraic polynomial under the square root in the denominator,
while the lower indice "$n$" denotes the order of the elliptic integral,
related to the "$y^{n}$" or "$x^{n}$" terms in the nominator of Eq. (\ref%
{II.11A}) and in the denominator of Eq. (\ref{II.11B}) (for $n$-th order
elliptic integrals of the third kind).

\subsection{\label{sec:Comparisons}Comparison with some definitions from
standard textbooks}

In the textbook \cite{Liashko} a statement is expressed (although without
any proof) that "higher-order elliptic integrals in the Legendre form can be
expressed by means of the three standard integrals
\begin{equation}
\int \frac{dy}{\sqrt{(1-y^{2})(1-q^{2}y^{2})}}\text{ \ \ , \ \ }\int \frac{%
y^{2}dy}{\sqrt{(1-y^{2})(1-q^{2}y^{2})}}\text{\ \ , \ }\int \frac{dy}{%
(1+hy^{2})\sqrt{(1-y^{2})(1-q^{2}y^{2})}},  \label{II.18}
\end{equation}%
where the parameter $h$ can be also a complex number", and these integrals,
as stated, "cannot be expressed through elementary functions".\ This
assertion is not precise, because in the preceding sections an example was
found, when the second integral in Eq. (\ref{II.18}) in fact can be
expressed through elementary function - for this case, by the logarithmic
term in Eq. (\ref{K11}). A similar statement to the one in \ \cite{Liashko}
is given also in the monograph \cite{Fichten2}: "Integrals of the type Eq. (%
\ref{II.1}) $\int R(y,\sqrt{G(y)})dy$ \ and $\ \int R(x,\sqrt{P(x)})dx$\ \
cannot be expressed by elementary functions in their final form, even if an
extended understanding of this notion is considered."

In the monograph \cite{PrasSol} a statement is expressed in a correct and
precise manner, namely: "Every elliptic integral can be represented in the
form of a linear combination of a rational function of the variables $y$ and
$\widetilde{y}$, integrals of a rational function of the variable $y$ and
also the integrals
\begin{equation}
\int \frac{dy}{\sqrt{G(y)}}\text{ \ , \ }\int \frac{ydy}{\sqrt{G(y)}}\text{
\ , }\int \frac{y^{2}dy}{\sqrt{G(y)}}\text{ \ , \ }\int \frac{dy}{(y-c)\sqrt{%
G(y)}}\text{ \ \ \ ."}  \label{II.20}
\end{equation}

In this statement it is not mentioned that elliptic integrals cannot be
expressed by elementary functions.

\section{\label{sec:SecCorrSpaceOrbits}Fourth-order elliptic integrals and
the second $O(\frac{1}{c^{3}})$ correction for the propagation time for the
case of a space-distributed orbit}

Taking into account the general equality Eq. (\ref{K5}), the second $O(\frac{1%
}{c^{3}})$ correction can be written as
\begin{equation}
T_{2}=-\int \frac{v}{c}.\frac{2V}{c^{2}}dt=-\frac{2G_{\oplus }M_{\oplus }}{%
c^{3}}\int \frac{v_{f}}{r}df\text{ \ \ ,}  \label{II.21}
\end{equation}%
where $v_{f}$, the radius-vector $r$ in the orbital plane and the potential $%
V$ of the gravitational field around the Earth are defined in the usual way
\begin{equation}
v_{f}=\frac{na}{\sqrt{1-e^{2}}}\sqrt{1+2e\cos f+e^{2}}\text{ \ \ , \ \ }r=%
\frac{a(1-e^{2})}{1+e\cos f}\text{ , \ \ \ }V=\frac{GM}{r}=\frac{GM}{%
a(1-e^{2})}(1+e\cos f)\text{\ .}  \label{II.22}
\end{equation}%
The final integral for the second correction is of the form
\begin{equation}
T_{2}=-\frac{2G_{\oplus }M_{\oplus }}{c^{3}}.\frac{na}{a\sqrt{1-e^{2}}}\int
(1+e\cos f)\sqrt{1+2e\cos f+e^{2}}df=T_{2}^{(1)}+T_{2}^{(2)}\text{ \ , }
\label{II.23}
\end{equation}%
where
\begin{equation}
T_{2}^{(1)}=-\frac{2G_{\oplus }M_{\oplus }}{c^{3}}.\frac{n}{(1-e^{2})^{\frac{%
3}{2}}}\int \sqrt{1+2e\cos f+e^{2}}df=-\frac{2G_{\oplus }M_{\oplus }}{c^{3}}.%
\frac{n}{(1-e^{2})^{\frac{3}{2}}}\widetilde{T}_{1}\text{ \ \ \ }
\label{II.24}
\end{equation}%
and $\ \widetilde{T}_{1}$ is the previously calculated integral $\
\widetilde{T}_{1}=\int \sqrt{1+2e\cos f+e^{2}}df$ . The second term $%
T_{2}^{(2)}$ in the second correction $T_{2}$ is \
\begin{equation}
T_{2}^{(2)}=-\frac{2G_{\oplus }M_{\oplus }}{c^{3}}.\frac{ne}{(1-e^{2})^{%
\frac{3}{2}}}\int \cos f\sqrt{1+2e\cos f+e^{2}}df\text{ }=-\frac{2G_{\oplus
}M_{\oplus }}{c^{3}}.\frac{ne}{(1-e^{2})^{\frac{3}{2}}}\widetilde{T}%
_{2}^{(2)}\text{\ \ ,}  \label{II.25}
\end{equation}%
where $\widetilde{T}_{2}^{(2)}$ is the notation for the more complicated
integral
\begin{equation}
\widetilde{T}_{2}^{(2)}=\int \cos f\sqrt{1+2e\cos f+e^{2}}df\text{ \ .}
\label{II.26}
\end{equation}

\subsection{The second part of the second $O(\frac{1}{c^{3}})$ \ correction
expressed in terms of elliptic integrals of the second and fourth order}

\bigskip It can be proved that%
\begin{equation}
T_{2}^{(2)}=-\frac{2G_{\oplus }M_{\oplus }}{c^{3}}.\frac{ne}{(1-e^{2})^{%
\frac{3}{2}}}\widetilde{T}_{2}^{(2)}\text{ }=  \label{II.27}
\end{equation}%
\begin{equation}
=-inq^{\frac{5}{2}}\frac{2G_{\oplus }M_{\oplus }}{c^{3}}(1+e^{2})\widetilde{J%
}_{2}^{(4)}(\widetilde{y},q)+inq^{\frac{3}{2}}\frac{2G_{\oplus }M_{\oplus }}{%
c^{3}}\frac{(1+e^{2})}{(1-e^{2})}\text{ }\widetilde{J}_{4}^{(4)}(\widetilde{y%
},q)\text{\ \ ,}  \label{II.28}
\end{equation}%
where
\begin{equation}
\widetilde{J}_{2}^{(4)}(\widetilde{y},q)=\int \frac{\widetilde{y}^{2}d%
\widetilde{y}}{\sqrt{\left( 1-\widetilde{y}^{2}\right) \left( 1-q^{2}%
\widetilde{y}^{2}\right) }}=\frac{1}{i}\int \frac{q^{3}\widehat{y}^{2}d%
\widehat{y}}{\sqrt{\left( \widehat{y}^{2}-1\right) \left( 1-q^{2}\widehat{y}%
^{2}\right) }}=\frac{q^{3}}{i}\text{ }\widetilde{J}_{2}^{(4)}(\widehat{y},q)%
\text{\ \ }  \label{II.29}
\end{equation}%
and $\widehat{y}$ is the notation for the variable $\widehat{y}=\frac{%
\widetilde{y}}{q}$. The last formulae can also be compared to Eq. (\ref{K21}%
). Analogously, the formulae for the fourth-order elliptic integral $%
\widetilde{J}_{4}^{(4)}(\widetilde{y},q)$ can be written as $\widetilde{J}%
_{4}^{(4)}(\widetilde{y},q)=\frac{q^{5}}{i}\int \frac{\widehat{y}^{4}d%
\widehat{y}}{\sqrt{\left( \widehat{y}^{2}-1\right) \left( 1-q^{2}\widehat{y}%
^{2}\right) }}$.

\subsection{Relation between the fourth-order and the second-order elliptic
integrals in the expression for $T_{2}^{(2)}$}

It can very easily be proved that
\begin{equation}
\int \sqrt{\left( 1-\widetilde{y}^{2}\right) \left( 1-q^{2}\widetilde{y}%
^{2}\right) }d\widetilde{y}=\frac{2}{3}\widetilde{J}_{0}^{(4)}-\frac{1}{3}%
(1+q^{2})\widetilde{J}_{2}^{(4)}+\frac{1}{3}\left[ \widetilde{y}\sqrt{\left(
1-\widetilde{y}^{2}\right) \left( 1-q^{2}\widetilde{y}^{2}\right) }\right]
\mid _{\widetilde{y}=\widetilde{y}_{0}}^{\widetilde{y}=\widetilde{y}_{1}}.
\label{II.30}
\end{equation}%
We have already proved that $\widetilde{J}_{2}^{(4)}$ can be expressed in
elementary functions, but yet we do not know whether $\widetilde{J}%
_{0}^{(4)} $ or the second-rank integral $\int \sqrt{\left( 1-\widetilde{y}%
^{2}\right) \left( 1-q^{2}\widetilde{y}^{2}\right) }d\widetilde{y}$ can also
be expressed through elementary functions. However, this will be proved in a
forthcoming publication.

Further, calculating the derivatives $\frac{d}{d\widetilde{y}}\left( \sqrt{%
\left( 1-\widetilde{y}^{2}\right) \left( 1-q^{2}\widetilde{y}^{2}\right) }%
\right) $ and $\frac{d}{d\widetilde{y}}\left( \widetilde{y}\sqrt{\left( 1-%
\widetilde{y}^{2}\right) \left( 1-q^{2}\widetilde{y}^{2}\right) }\right) $,
integrating the resulting equations from $\widetilde{y}_{0}$ to $\widetilde{y%
}_{1}$, and combining all the three equations, the following two equations
can be derived for $\widetilde{J}_{4}^{(4)}$ and \ $\widetilde{J}_{3}^{(4)}$%
:
\begin{equation}
\widetilde{J}_{4}^{(4)}=\frac{1}{3q^{2}}\left[ \widetilde{y}\sqrt{\left( 1-%
\widetilde{y}^{2}\right) \left( 1-q^{2}\widetilde{y}^{2}\right) }\right]
\mid _{\widetilde{y}=\widetilde{y}_{0}}^{\widetilde{y}=\widetilde{y}_{1}}+%
\frac{2(1+q^{2})}{3q^{2}}\widetilde{J}_{2}^{(4)}-\frac{1}{3q^{2}}\widetilde{J%
}_{0}^{(4)}\text{ \ \ ,}  \label{II.31}
\end{equation}%
\begin{equation}
\widetilde{J}_{3}^{(4)}=\frac{1}{2q^{2}}\left[ \sqrt{\left( 1-\widetilde{y}%
^{2}\right) \left( 1-q^{2}\widetilde{y}^{2}\right) }\right] \mid _{%
\widetilde{y}=\widetilde{y}_{0}}^{\widetilde{y}=\widetilde{y}_{1}}+\frac{%
(1+q^{2})}{2q^{2}}\widetilde{J}_{1}^{(4)}\text{ \ \ \ .}  \label{II.32}
\end{equation}%
In expressions Eq. (\ref{II.30}) - Eq. (\ref{II.32}) the symbol "$\mid _{%
\widetilde{y}=\widetilde{y}_{0}}^{\widetilde{y}=\widetilde{y}_{1}}$" means
that the corresponding expression is taken at $\widetilde{y}=\widetilde{y}%
_{1}$ and from it the value of the expression at $\widetilde{y}=\widetilde{y}%
_{0}$ is substracted.

We should also add to this system of equations the earlier derived equation
for $\widetilde{J}_{2}^{(4)}$\
\begin{equation}
\widetilde{J}_{2}^{(4)}(\widetilde{y},q)=\int \frac{\widetilde{y}^{2}d%
\widetilde{y}}{\sqrt{(1-\widetilde{y}^{2})(1-q^{2}\widetilde{y}^{2})}}=-%
\frac{1}{q^{2}}E(\varphi )+\frac{1}{q^{2}}\widetilde{J}_{0}^{(4)}(\widetilde{%
y}^{2};q)\text{ \ \ .}  \label{II.33}
\end{equation}%
\ The second equation Eq. (\ref{II.32}) for $\widetilde{J}_{3}^{(4)}$
clearly suggests that $\widetilde{J}_{3}^{(4)}$ is expressed in elementary
functions because the first-order elliptic integral $\widetilde{J}_{1}^{(4)}$
$=\int \frac{\widetilde{y}d\widetilde{y}}{\sqrt{(1-\widetilde{y}^{2})(1-q^{2}%
\widetilde{y}^{2})}}$ can be analytically calculated after performing the
Euler substitution
\begin{equation}
\sqrt{\widetilde{y}^{4}-\frac{(1+q^{2})}{q^{2}}\widetilde{y}^{2}+\frac{1}{%
q^{2}}}=u+\widetilde{y}^{2}\text{ \ \ .}  \label{II.34}
\end{equation}%
The result is
\begin{equation}
\widetilde{J}_{1}^{(4)}=-\frac{1}{2q^{3}}\ln \mid \sqrt{(1-\widetilde{y}%
^{2})(1-q^{2}\widetilde{y}^{2})}-\widetilde{y}^{2}+\frac{(1+q^{2})}{2q^{2}}%
\mid _{\widetilde{y}=\widetilde{y}_{0}}^{\widetilde{y}=\widetilde{y}_{1}}%
\text{ .}  \label{II.35}
\end{equation}

\bigskip Note that the variable $\widetilde{y}$ is not related to any
elliptic function! Concerning the fourth-order elliptic integral $\widetilde{%
J}_{4}^{(4)}$, taking into account the already proved fact that $\widetilde{J%
}_{2}^{(4)}$ can be expressed in elementary functions, it would follow that $%
\widetilde{J}_{4}^{(4)}$ can also be expressed by elementary functions, but
it is necessary to prove that the zero-order elliptic integral $\widetilde{J}%
_{0}^{(4)}$ can be expressed in elementary functions. In another
publication, it will be proved that the integral can be expressed by an
analytical formulae, but also it can be numerically calculated, since the
modulus of the elliptic integral is equal to the eccentricity $%
e=0.01323881349526$ of the GPS orbit, which is exactly known \cite%
{Gulklett:C4}. The general mathematical theory for the recurrent relations
between $n-$th order elliptic integrals of any kind and the lower-order
elliptic integrals has been developed in the monographs \cite{PrasSol}, \cite%
{Smirnov3} and \cite{Fichten2}.

\section{Conclusion}

In this paper we considered two cases of propagation of a signal in the
gravitational field around the Earth, when General Relativity effects have
to be taken into account: 1 case. Signals, exchanged between satellites on
one and the same plane elliptical orbit; 2 case. Signals, exchanged between
satellites on a space-distributed orbit. It has been shown that in the first
case the propagation time is given by a sum of elliptic integrals of first,
second and the third kind, all of which however are of zero-order. The
integrals are represented in the Legendre form. In the second case, the
propagation time is expressed by a more complicated combination of elliptic
integrals of the second- and of the fourth- order. It is remarkable also
that the fourth-order elliptic integral can be decomposed into elliptic
integrals of the second kind and of the first kind, which are characteristic
for the first case. Yet, an elliptic integral of the third kind appears in
the first case and does not appear in the second case. All the relevant
integrals in the expressions for the propagation time have coefficients,
which have definite numerical values, so from the found analytical
expressions the numerical value for the propagation time can also be calculated. 
In principle, elliptic integrals of zero
order and of the first-, second- and third-  kind can be evaluated also
numerically.

The method of calculation in both cases is based on the null cone formalism
in General Relativity, which previously has been applied in many other
papers. This method is a consequence of the light-like geodesics formalism
in \ GR, so it is a reliable one. However, in this aspect some theoretical
problems yet remain not investigated- for example, if light-like geodesics
are applied, will they give some new solution about the propagation time.
Such an investigation is very complicated from a mathematical point of view,
because the system of equations for the null geodesics is a nonlinear one
and probably will require methods from the theory of dynamical systems and
not the iterative approximation methods, applied for the moment. An inconvenience of 
the proposed in this paper method is that one should know the position of the signal-receiving 
satellite at the moment of reception of the signal. This position depends on the 
time of motion of the satellite, but since it is equal to the propagation time, 
it is also an unknown variable. This problem can be solved if the formalism of 
two null cones, developed in \cite{BOGDAN:PRD} and \cite{BOGDAN:AIP2019} is applied. 
In fact, this inconvenience is typical also for the Shapiro delay formulae. 

Another problem which for the moment is not yet solved \ is about finding
the propagation time for signals, exchanged between satellites on different
space-distributed orbits, which is important from an experimental point of
view in reference to the NASA and ESA concept about autonomous navigation.
Since the Keplerian parameters of the two orbits are different, one might
propose again to use the null cone formalism, but in some way it
might be thought that the orbit is  "gradually deformed" from the first set of $\ $%
Keplerian parameters $(f_{1},a_{1},e_{1},\Omega _{1},i_{1},\omega _{1})$ to
the second set of parameters$\ (f_{2},a_{2},e_{2},\Omega _{2},i_{2},\omega
_{2})$. This means that in the null-cone equation not only the true anomaly
angle $f$ will change, but also all the other parameters of the
space-distributed orbit, and integration has to be performed over all these
parameters. This will represent a considerable mathematical difficulty. 

\begin{acknowledgments}
The author is grateful to the organizers of the Thirteenth Hybrid Conference of the
Euro-American Consortium for Promoting  the Application of Mathematics in Technical and
Natural Sciences   (AMITaNS’21) 24-29 June 2021, Albena resourt, Bulgaria and especially to
Prof. Michail Todorov for the opportunity to participate in this conference.
\dots.
\end{acknowledgments}


\end{document}